\documentclass[twocolumn]{aastex631}

\newcommand\binfrac{0.41}
\newcommand\binfracsig{0.02}

\newcommand\binaries{571 }
\newcommand\trh{230 }

\usepackage{graphicx}
\usepackage{lipsum}
\usepackage{float}
\begin{document}

\title{Tracing the Origins of Mass Segregation in M35: Evidence for Primordially Segregated Binaries}

\author[0009-0001-9841-0846]{Erin Motherway}
\affiliation{Embry-Riddle Aeronautical University, Department of Physical Sciences, 1 Aerospace Blvd, Daytona Beach, FL 32114, USA}

\author[0000-0002-3881-9332]{Aaron M. Geller}
\affiliation{Center for Interdisciplinary Exploration and Research in Astrophysics (CIERA) and Department of Physics and Astronomy, Northwestern University, 1800 Sherman Avenue, Evanston, IL 60201, USA}

\author[0000-0002-9343-8612]{Anna C. Childs}
\affiliation{Center for Interdisciplinary Exploration and Research in Astrophysics (CIERA) and Department of Physics and Astronomy, Northwestern University, 1800 Sherman Avenue, Evanston, IL 60201, USA}

\author[0000-0003-3695-2655]{Claire Zwicker}
\affiliation{Illinois Institute of Technology, 10 West 35th Street Chicago, IL 60616, United States of America}

\author[0000-0002-5775-2866]{Ted von Hippel}
\affiliation{Embry-Riddle Aeronautical University, Department of Physical Sciences, 1 Aerospace Blvd, Daytona Beach, FL 32114, USA}

\begin{abstract}

M35 is a young open cluster and home to an extensive binary population. 
Using Gaia DR3, Pan-STARRS, and 2MASS photometry with the Bayesian statistical software, BASE-9, we derive precise cluster parameters, identify single and binary cluster members, and extract their masses. We identify \binaries\ binaries down to Gaia $G = 20.3$ and a lower-limit on the binary frequency of $f_b = \binfrac \pm \binfracsig$. We extend the binary demographics by many magnitudes faint-ward of previous (radial-velocity) studies of this cluster and further away from the cluster center ($1.78^{\circ}$, roughly 10 core radii). We find the binary stars to be more centrally concentrated than the single stars in the cluster.  Furthermore, we find strong evidence for mass segregation within the binary population itself, with progressively more massive binary samples becoming more and more centrally concentrated. For the single stars, we find weaker evidence for mass segregation; only the most massive single stars ($>2.5M_\odot$) appear more centrally concentrated. Given the cluster age of $\sim\ $200~Myr, and our derived half-mass relaxation time for the cluster of $\trh \pm 84$~Myr, we estimate $\sim47$\% of the binary stars and $\sim12$\% of single stars in the cluster have had time to become dynamically mass segregated. Importantly, when we investigate only stars with mass segregation timescales greater than the cluster age, we still find the binaries to be more centrally concentrated than the singles, suggesting the binaries may have formed with a primordially different spatial distribution than the single stars. 

\end{abstract}


\section{Introduction} \label{sec:intro}

Open clusters (OCs) provide an important laboratory for studying dynamical evolution and mass segregation, given their typically extensive amount of stellar multiplicity and the homogeneity of stellar properties such as age, distance, metallicity, reddening, etc. in a given cluster. 
Through two-body relaxation and other processes leading to energy exchange between stars and binaries, star clusters are expected to become (dynamically) mass segregated, where the more massive objects in the cluster occupy a more centrally concentrated spatial distribution than the lighter objects \citep{1962AJ.....67..471K, 1997MNRAS.286..709G}.  For a given mass range in (primary) stars, binaries are expected to be more massive than the single stars (since binaries contain two stars), and therefore the expectation for relaxed clusters is that binaries will be mass segregated with respect to single stars.  Furthermore, binary systems offer significant insight into the dynamical evolution of OCs as most gravitational encounters in OCs are expected to be influenced by binaries, and can lead to direct collisions, mergers, mass transfer, and energy exchanges \citep[e.g.,][]{2004MNRAS.352....1F, 2011MNRAS.410.2370L, 2017IAUS..316..222D}. 

The dynamical states and evolution of young clusters are of particular interest, as they are expected to retain their near-birth (primordial) conditions. Searching for mass segregation in the youngest clusters can address the question of whether mass segregation may be primordial or, largely attributed to dynamical and relaxation processes. Some of the youngest clusters that have been studied observationally for mass segregation include  NGC 2264  \citep[3 Myr;][]{2021A&A...645A..94N}, NGC 6231  \citep[3-4 Myr;][]{1998A&A...333..897R},  IC 1590  \citep[3.5–4.4 Myr;][]{2021AJ....162..140K},  NGC 2516  \citep[100 Myr;][]{2022BAAA...63..121P} and the Hyades  \citep[680 Myr;][]{2022MNRAS.512.3846E},. 

NGC 2264, NGC 6231, IC 1590, and NGC 2516 all show evidence of primordial mass segregation.  Each of these clusters also have relaxation times that are $\gtrsim$ the cluster age.  In particular for NGC 6231, \citet{1998A&A...333..897R} concluded that binaries (and other multiple systems) are more centrally concentrated than single stars, likely as a result of stellar formation processes and not through dynamical means.
In contrast, \citet{1998A&A...333..897R} find the Hyades to contain a highly mass segregated stellar population as a result of dynamical relaxation. This may be due to the Hyades having already undergone $\sim 10$ relaxation times.
At an age of $\sim\ $130-200 Myrs  \citep{2010AJ....139.1383G, 1986ApJ...310..613M, 2015AJ....150...10L, 2008IAUS..246..105B, 2011AJ....142...53M}, M35 lies in between these previously studied clusters. 

Numerical simulations of similar young clusters have also been used to study the origin of mass segregation. \citet{1998MNRAS.295..691B} found within their simulations that the most massive members of a young OC most likely formed in the center of that cluster. \citet{2020A&A...638A.155P} concluded that binaries lead to quicker dynamical evolution for an OC like the Orion Nebula Cluster (ONC). Clusters with initially larger binary fractions evolve more quickly due to the influence of massive binaries (and even more so by hard binaries). The study also finds that the ONC was likely primordially mass segregated. \citet{2010MNRAS.405..666C} determined through a simulation that the Pleiades was also initially mass segregated.

M35, is another young and well studied OC that is ripe for studies of mass segregation.  The cluster has multiple photometric \citep{2002AJ....124.1555V,2009AAS...21340708A,2003AJ....126.1402K} and  kinematic \citep{2011AJ....142...53M,2010AJ....139.1383G,1983PhDT.........8M} studies that show M35 has a rich stellar population including many binary stars. Furthermore, previous studies found that though M35 is not yet completely relaxed, only the more massive members appear to show some degree of mass segregation \citep{2001ApJ...546.1006B,1986ApJ...310..613M,2008IAUS..246..105B}.  Thus, M35 offers an interesting test case for investigating the interplay between primordial and dynamical mass segregation.

In this paper, we use high precision photometry and kinematic constraints from Gaia DR3 to extend the analysis of mass segregation fainter (9 $< G <$ 20) and further out in radius from the cluster center (to a $1.78^{\circ}$ radius, or roughly 10 core radii) than previous studies using a new method of identifying photometric binary stars and their masses. We focus on mass segregation of the binaries in M35, and whether we find evidence for primordial mass segregation.  In Section~\ref{sec:methods}, we describe our methods and  our stellar sample. In Section~\ref{sec:results}, we investigate our data for mass segregation and provide relaxation timescales for M35. In Section~\ref{sec:discussion}, we discuss the origin of the mass segregation detected in M35, and in Section~\ref{sec:conclusions} we summarize our findings.

\section{Defining the stellar population of M35}\label{sec:methods}

We use the Bayesian Analysis of Stellar Evolution with Nine Parameters (BASE-9) software suite to derive global cluster parameters and star-by-star characteristics for M35 cluster members \citep{2006ApJ...645.1436V, 2009AnApS...3..117V, 2016ascl.soft08007R}. BASE-9 uses input photometry and stellar evolution models (we use PARSEC isochrone models for this study \citealt{2012MNRAS.427..127B}) with a Markov chain Monte Carlo (MCMC) technique to estimate the posterior probability distribution for global cluster properties including age, metallicity, distance modulus, and line-of-sight absorption as well as star-by-star properties of cluster (photometric) membership, stellar masses and mass ratios for likely binary stars. See \citet{2020AJ....159...11C} for more information.

To generate our input stellar sample for BASE-9, we consider all stars in the Gaia database within an effective radius of $60'$. \citet{2023arXiv230816282C} estimate this effective radius by eye by fitting a \citet{King1962} model to the observed radial surface density distribution and identifying where the observations depart from the model.

\begin{figure}[!t]
    \includegraphics[width=0.45\textwidth]{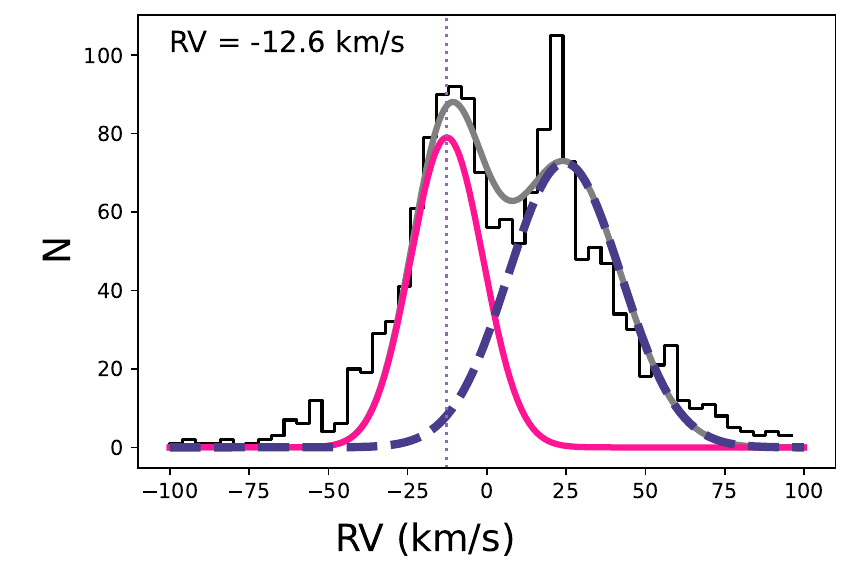}
    \includegraphics[width=0.45\textwidth]{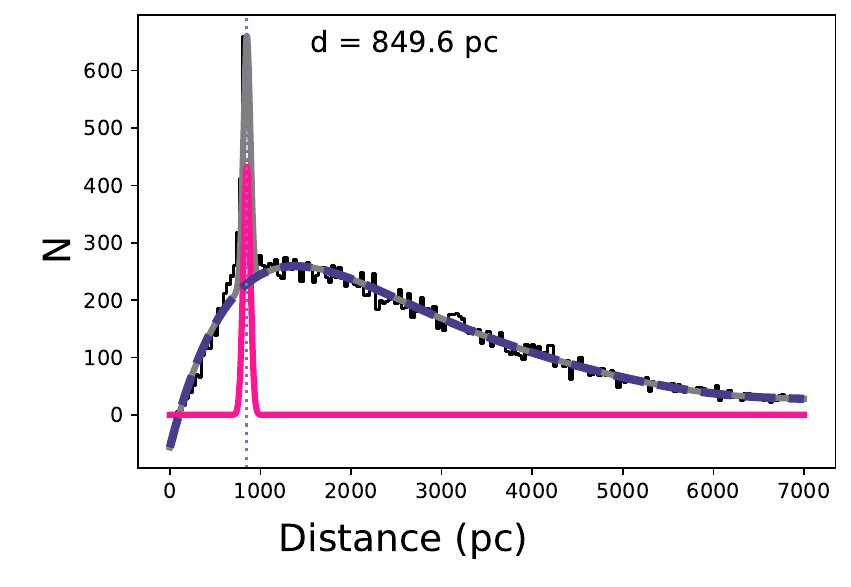}
    \includegraphics[width=0.45\textwidth]{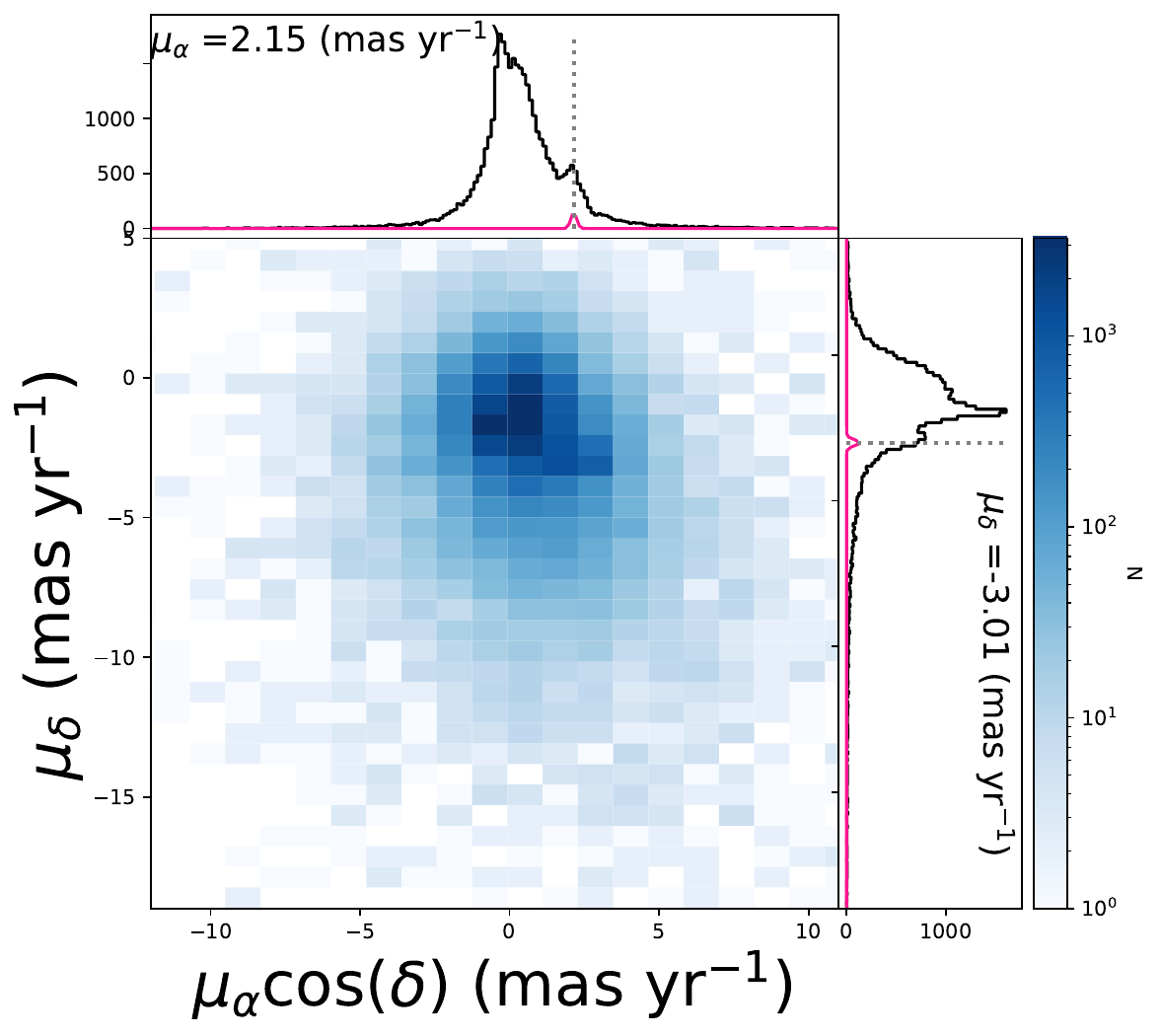}
    \caption{Gaussian fits for M35 for stars within a $~0.5^{\circ}$ radius from the cluster center. These fits for the RV (top), distance (middle), and PM (bottom) distributions are shown in pink, where vertical dotted lines represent the mean of the Gaussian fit. Dashed purple lines indicate the Gaussian fit to the field RV data (top) and a sixth order polynomial fit to the field star distances (middle). Combined fits are represented by gray curves in the top and middle plots.
     \label{fig:fig1}}
\end{figure}

 In order to begin separating field stars from cluster members within this sample, we fit Gaussian functions to the cluster distributions in  observed radial-velocity (RV), distance and proper-motion (PM) (all from Gaia). For the RV distribution we also fit a second Gaussian simultaneously to the field distribution, and for the distance distribution we simultaneously fit a polynomial function to the field distribution.  (We have found that fitting multiple functions to the PM distribution does not improve our analysis.) These fits for M35 are shown in Figure \ref{fig:fig1}. We then use the cluster Gaussian fits to estimate a membership probability for each star, based on their distance from the mean of each respective Gaussian.  We exclude from our sample any star that is $>10\sigma$ from the mean of the respective cluster Gaussian, in any dimension, as a very conservative cut on membership (which will be further refined by the photometric analysis in BASE-9). All underlying data for this paper has been published to Zenodo (doi:10.5281/zenodo.10080762). Please see \citet{2023arXiv230816282C} for more details.  
  
 We then remove the differential component of the reddening from the Gaia $G$, $G_{BP}$, $G_{RP}$ \citep{2023A&A...674A...1G}, Pan-STARRS $g$, $r$, $i$, $z$, $y$ \citep{2020ApJS..251....6M}, and 2MASS $J$, $H$, $K$ \citep{2006AJ....131.1163S} photometry, using the \texttt{Bayestar19} model \citep{Green2019} as described in \citet{2023arXiv230816282C}, and feed this reddening corrected photometry for these potential cluster members through BASE-9. We  provide our calculated membership probabilities for each star as priors.  BASE-9 then identifies likely photometric members based on each star's distance from the isochrone model in all filters.

We then impose that the cluster membership probabilities from both Gaia and BASE-9 are each greater than zero.  As a further step to limit field star contamination, we only include stars that have an uncertainty in the Gaia $G_{BP}$  filter  less than the median uncertainty for this filter over the full sample, of 0.13 mag. We choose to use the Gaia $G_{BP}$  filter because it has the highest median uncertainty value over all other Gaia and Pan-STARRS filters. We also choose to limit our analysis to only include main sequence stars, and therefore we remove any giants (Gaia $G <$ 9) from our sample. (We do, however, include the masses of these excluded giants when determining the total mass of the cluster, which is used to estimate relaxation timescales in Section~\ref{sec:results}.) We use all stars that meet these requirements as our final sample for the subsequent analysis. 

We also note that only the brighter stars in our sample have RV measurements, and therefore our membership selection differs slightly between the bright and faint samples. This also offers us an opportunity to estimate an amount of field star contamination in our sample. Of the 12844 stars in our initial sample that have Gaia RV measurements, we find 281 to be (Gaia) members of M35 when considering RV, PM and distance.  If we ignore RV, we find 30 additional potential M35 members in that sample (of 12844 stars).  After running these 30 stars through BASE-9 we find that 26 are considered cluster members based on their photometry.  Matching to the WIYN Open Cluster Study (WOCS) RV survey for M35  \citep{2010AJ....139.1383G, 2015AJ....150...10L} and checking the Gaia RUWE value suggests that 18/26 of these stars are likely binaries; some of these may indeed be cluster members where the Gaia RV does not represent the system center of mass (and therefore the measured Gaia RV is outside the cluster distribution).  If we consider the remaining 8/26 of these stars to in fact be field stars, this would suggest that the cluster members without RV measurements may have an additional field star contamination level of  $8/(281+30) \sim 2.5$\%.

As a further check on our cluster members, we compare to the recent membership analysis from WOCS \citep{2010AJ....139.1383G, 2015AJ....150...10L}; we recover 413/418 (99\%) of their members (including their single members, binary members, and binary likely members).  

\begin{figure}[!t]
    \plotone{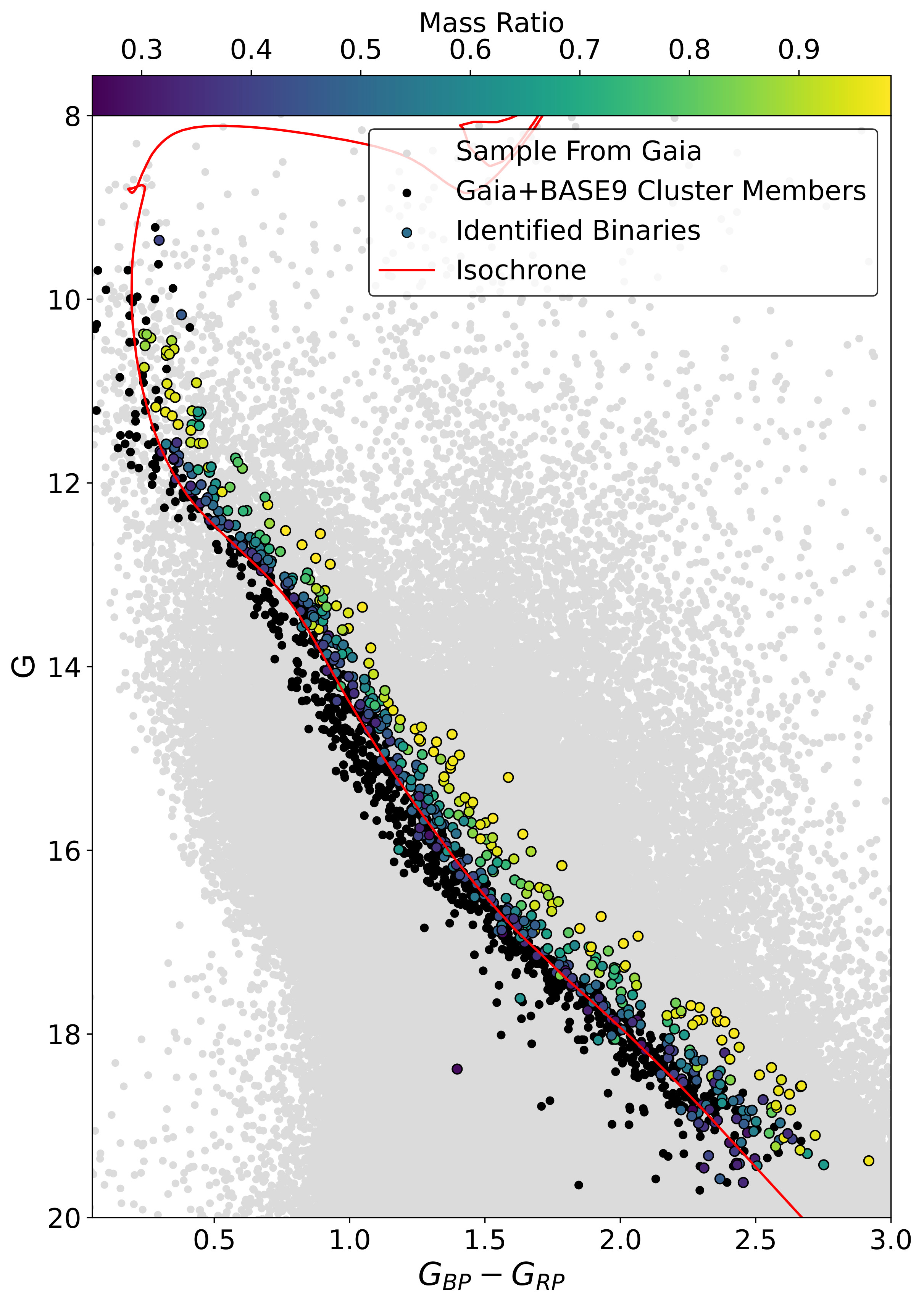}
    \caption{Color-Magnitude Diagram of M35, including the entire sample we  downloaded from the Gaia archive (light gray), BASE-9 identified cluster members (overplotted in black), and identified binaries (overplotted in colored symbols, indicating the mass ratio). With the red line, we show a PARSEC isochrone defined by the median cluster parameters found by BASE-9.
    \label{fig:fig0}}
\end{figure}

Finally, BASE-9 also provides posterior distributions of primary masses and mass ratios for each star system. For a given stellar system, we require the median value of the posterior distribution of its secondary mass to be $\geq3\sigma$ above zero to be considered a binary (following the procedure of \citealt{2020AJ....159...11C}). Our final sample, including our selected binaries and their corresponding mass ratios is displayed in Figure \ref{fig:fig0}. Note that in this color-magnitude diagram, we utilize one combination of filters from Gaia, but BASE-9 uses all filter combinations and their uncertainties to characterize a cluster. Please see \citet{2023arXiv230816282C} for a more detailed methodology.

\section{Mass Segregation in M35}\label{sec:results}
\subsection{Mass Segregation Signatures}

We first investigate the binary fraction of the cluster.  In our final sample of cluster members, we find a binary fraction of $\binfrac \pm \binfracsig$. This is a lower limit on the true binary fraction because our method becomes incomplete for binaries with very low mass ratios \citep{2020AJ....159...11C}. For comparison, \citet{2010AJ....139.1383G} found a radial-velocity binary fraction of $0.24 \pm 0.03$ out to orbital periods of 10$^4$ days (over all mass ratios, but only for the brighter stars in our sample).  The larger binary fraction in our results is likely due both to the inclusion of longer-period binaries and brighter (more massive) stars that were inaccessible to the \citet{2010AJ....139.1383G} radial-velocity survey.

\begin{figure}[!t]
    \plotone{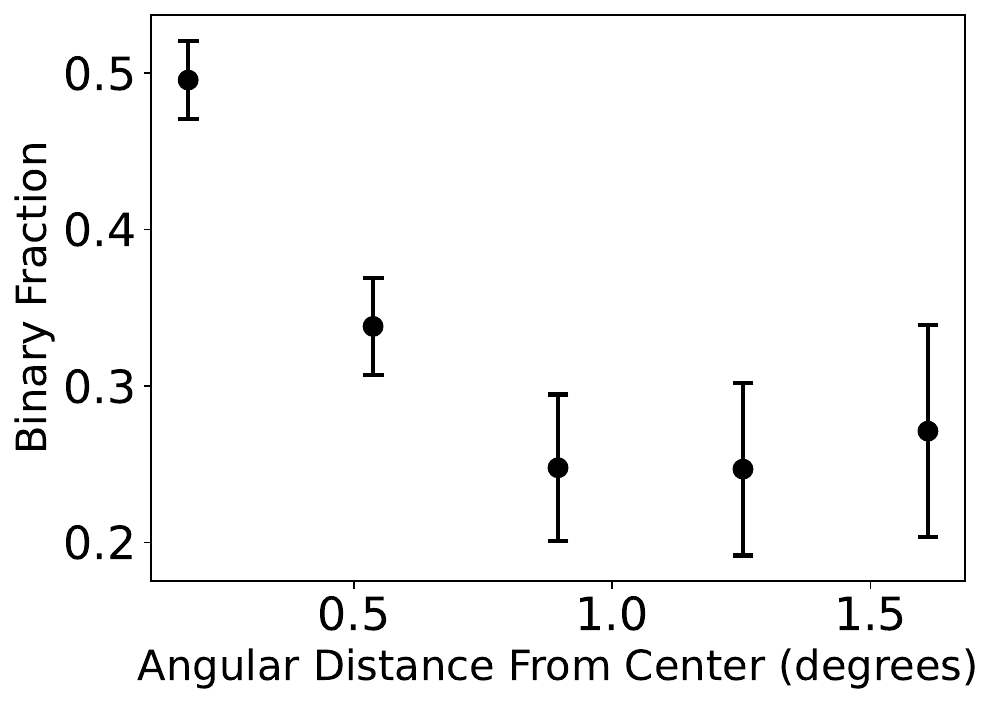}
    \caption{Binary fraction with respect to distance from the cluster center. We plot the binary fraction for five equal-sized bins within $1.75^{\circ}$ from the cluster center. The binary fraction is calculated using all binaries and single stars in our sample, but is not corrected for incompleteness. 
    \label{fig:fig2}}
\end{figure}

We also investigate the binary fraction as a function of radius from the cluster center (Figure \ref{fig:fig2}). The binary fraction is highest near the cluster center and drops off with increasing distance. Outside of $\sim0.8^{\circ}$ from the cluster center, the binary fraction plateaus at $\sim\ $25\%. A two-sided Z-test between the first (where $f_b$ is highest) and the last three (where $f_b$ plateaus) data points in Figure~\ref{fig:fig2} results in a p-value of $8.71\times 10^{-6}$.  Therefore, we conclude that the binaries are centrally concentrated with respect to the single stars in the cluster. 
 
\begin{figure*}
    \centering
    \includegraphics[width=0.95\textwidth]{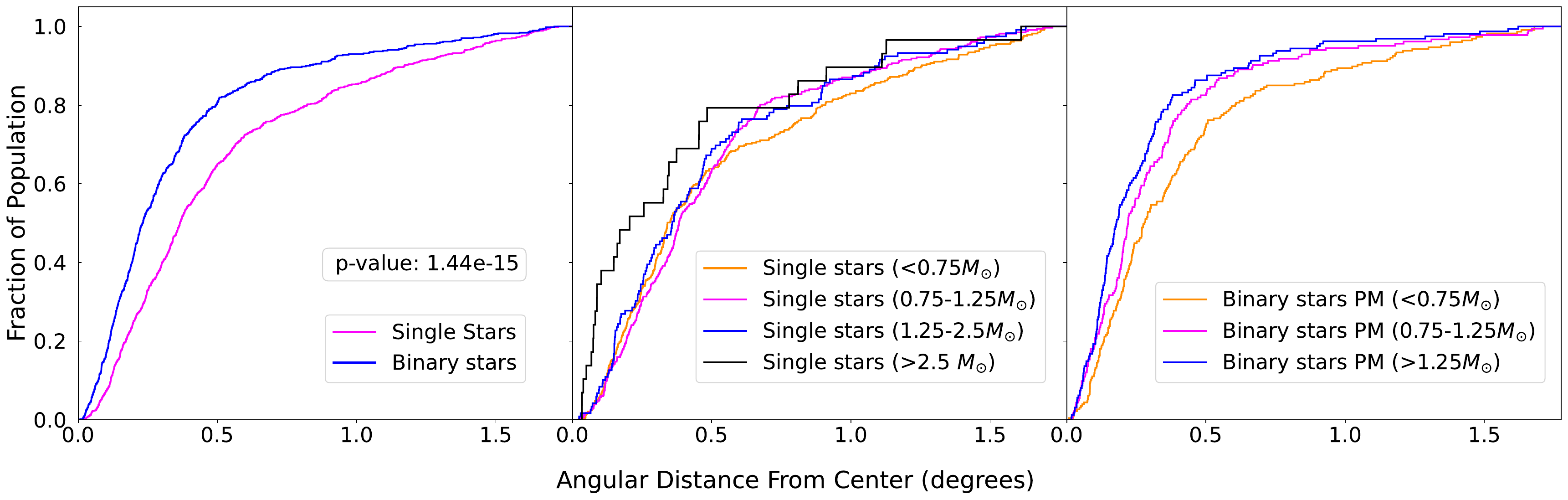}
    \caption{Cumulative distribution functions of the (left) binary and single star population, (middle) different mass ranges of the single star sample, and (right) different mass ranges of the binary star sample.  (For binaries, we use the primary mass of the system.) We also show a p-value in the left panel resulting from a K-S test between the binary and single star samples. Resulting p-values from K-S tests between the highest and lowest mass ranges for the middle and right panel are $7.79\times 10^{-3}$ and $3.44\times 10^{-5}$, respectively.}
    \label{fig:fig3}
\end{figure*}

 More evidence for the central concentration of the binary stars of M35 is evident in Figure \ref{fig:fig3}. In the left panel, we plot a cumulative distribution function (CDF) of our entire single and binary star sample in M35. Visually, the figure shows that the binaries are more centrally concentrated as compared to the single stars.  A Kolmogorov-Smirnov (K-S) test of whether these populations were drawn from the same distribution results in a p-value of $1.44 \times 10^{-15}$.  The radial distributions of the binary and single stars are drawn from distinct parent populations at very high confidence.  

 The average mass of the single stars in our sample is $0.98$ M$_\odot$, while the average (total) mass of the binary stars in our sample is $1.69$ M$_\odot$.  Thus we interpret this central concentration of the binaries as evidence for mass segregation in the cluster.

We also investigate single stars independently for evidence of mass segregation. As seen in the middle panel of Figure \ref{fig:fig3}, dividing the single star population in bins of mass reveals a central concentration of only the most massive objects. A K-S test comparing low-mass single stars to the high-mass single stars returns a p-value of $7.79 \times 10^{-3}$.  Similar K-S tests comparing the radial distributions of the single stars in the other mass bins does not return any significant distinction.

On the other hand, when we analyze the binary sample in different mass ranges (right panel of Figure \ref{fig:fig3}), mass segregation is plainly evident. The p-value of a K-S test comparing the binaries in the highest-mass bin to those in the lowest mass bin is $3.44 \times 10^{-5}$, indicating a significant difference in the respective radial distributions. Similar K-S tests comparing the radial distribution of the binary stars in the other mass bins all return p-values less than 0.03. Thus we find that the binary population displays strong evidence of mass segregation in M35. 

\subsection{Timescales}

To help interpret these results, we calculate various timescales for the cluster.  The half-mass relaxation time of the cluster is given by  
\begin{equation}
\label{eq1}
t_{rh}=\frac{0.17N}{\ln(\lambda N)}\sqrt{\frac{r_h^3}{GM}}
\end{equation}
where $N$ is number of cluster members, $\lambda$ is a constant for which we use a value of 0.1 \citep{1994MNRAS.268..257G}, $r_h$ is the half-mass radius of the cluster, and $M$ is the total mass of the cluster \citep{2008gady.book.....B, 1969ApJ...158L.139S}.

By investigating the radial mass distribution of all cluster members in our sample we estimate a half-mass radius of $5.2 \pm 0.1$~pc (at a distance of 826 pc, and accounting for projection effects by multiplying our observed projected half-mass radius by a factor of 4/3, following \citealt{1987degc.book.....S})).  For the total cluster mass and number of stars we find $M\approx1800$ M$_\odot$ and $N\approx1400$.  \citet{1989ApJ...339..195L} and \citet{ 2001ApJ...546.1006B} also provide estimates for the cluster mass and number of stars, which are somewhat higher than our values.  For our calculation of $t_{rh}$ we use $M = 2400$ $\pm$ $700$ M$_\odot$ and $N=2050$ $\pm$ $650$, which lie roughly at the center of the available measurements (including ours and the literature values) with uncertainties that span the range in measurements. These values result in $t_{rh}=230$ $\pm$ $84$ Myr, in agreement with \citet{1983PhDT.........8M}, \citet{2003AJ....126.1402K} and also with the value we calculate using our data alone of $\sim$200 Myr.  

To calculate the minimum time for mass segregation based on the most massive member of the cluster, we follow \cite{1969ApJ...158L.139S}:
\begin{equation}
\label{eq2}
t_{seg}=\frac{<m>}{m}t_{rh}
\end{equation}
where $ \langle m \rangle$ is the average mass of a star in the cluster, $m$ is the mass of any chosen star, and $t_{rh}$ is the half-mass relaxation time for the cluster.
The most massive star we find in our sample has a mass of $5.02$ M$_\odot$. We estimate $t_{seg}(5.02$ M$_\odot$$)=51$ Myr, which means that the most massive stars in M35 would have ample time to achieve mass segregation already.

\begin{figure}[!t]
    \plotone{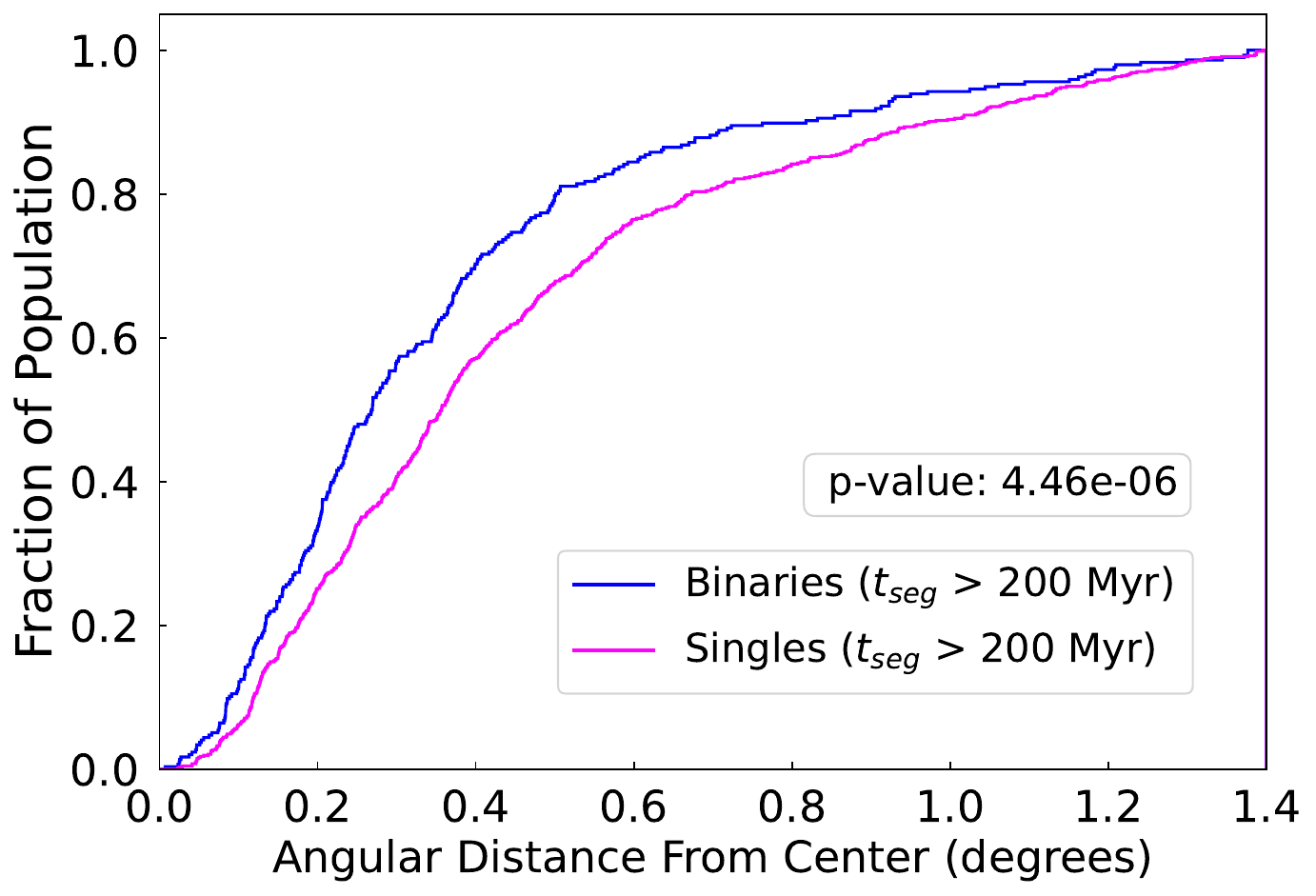}
    \caption{Cumulative distribution functions of the binary and single star population of M35 for stars with $t_{seg} > 200 $ Myr. The p-value from a K-S test comparing both populations is also shown.}
    \label{fig:fig4}
\end{figure}

Given the high quality of photometric data, we have sufficient low-mass (faint) cluster members to investigate a sample with predicted $t_{seg}$ greater than the cluster age, in order to probe for evidence of primordial mass segregation. $t_{seg}$ equals the cluster age of 200 Myr for an object of mass $1.28$ M$_\odot$, and our sample extends to a mass of $0.52$ M$_\odot$ ($0.39$ M$_\odot$) for binaries (singles).  In Figure~\ref{fig:fig4} we compare the radial distributions of the binary and single stars with  $t_{seg} > 200$ Myr and find the binaries to be centrally concentrated as compared to the single stars, at very high statistical significance (with a p-value of $4.46\times 10^{-6}$).  In this sample, the average mass of the binary (single) stars is $1.06$ M$_\odot$ ($0.82$ M$_\odot$).  

We also attempted to divide our single star and binary star samples, respectively into bins of $t_{seg} < 200 $ Myr and $t_{seg} > 200$ Myr, to compare their radial distributions.  We did not find any statistically significant distinction for the single or binary radial distributions in this test, though we note this may be due to small sample sizes, particularly for the binaries.   

\section{Discussion}\label{sec:discussion}

We detect a strong signature of mass segregation for the binary stars in M35. On average, the binary stars are more massive than the singles, and therefore have a shorter timescale for mass segregation. The time for mass segregation to occur in the M35 single stars ranges from $\sim$80-760 Myr with an average value of $\sim$360 Myr, while binary stars range from $\sim$60-570 Myr with an average of $\sim$220 Myr. Given the approximate cluster age of 200 Myr, we estimate 46\% of binary stars and 12\% of single stars have had time to become dynamically mass segregated. 

This may explain why similarly significant mass segregation is not seen for single stars (Figure \ref{fig:fig3}). Much of the single star population of M35 has not had enough time to undergo dynamical mass segregation.  Indeed we only find evidence of mass segregation for the most massive single stars (middle panel of Figure~\ref{fig:fig3}).

Previous studies of M35 by \citet{1983PhDT.........8M} and \citet{1986ApJ...310..613M} have shown the cluster to be mass segregated when dividing the sample in bins of magnitude, but without dividing the sample by singles and binaries. As a check, we replicate the bins from \citet{1983PhDT.........8M} and \citet{1986ApJ...310..613M}, though using our available photometric filters. If we divide the sample in bins of $G=$12, 14, 16, and 20 we do indeed find evidence for mass segregation between the brightest bin to the faintest bin (with a p-value of $3.4\times 10^{-5}$), in agreement with \citet{1983PhDT.........8M} and \citet{1986ApJ...310..613M}. However, if we only include the single stars in that analysis, we only see marginal indications of mass segregation (with a p-value of $2.4\times 10^{-2}$). We postulate that the mass segregation seen in these previous papers is coming (at least primarily) from the binaries.  

Importantly, we find the binaries to be more centrally concentrated than the single stars even for a sample with $t_{seg} > 200$ Myr. 
These stars are not expected to have had time to mass segregate through dynamics given the current cluster structure and number of stars (which was presumably larger at birth). If the cluster formed in a roughly spherical distribution without significant substructure, this suggests that the binaries \textit{formed} closer to the cluster center than single stars.  Alternatively, simulations suggest that this result could also be achieved via star formation in smaller clumps, that either formed mass segregated or had short enough relaxation times to quickly achieve mass segregation, and then merged to form the cluster observed today \citep{McMillan2007}. The details of gas expulsion from the young embedded cluster may also enhance early mass segregation \citep{Marks2008}. 
 Regardless of the specific mechanism, we interpret this as evidence for primordial, or near primordial, central concentration of the binaries during the formation process of M35. 

\section{Conclusions}\label{sec:conclusions}
Using Gaia DR3, Pan-STARRS, and 2MASS photometry with the Bayesian statistical software suite, BASE-9, we identify single and binary cluster members and extract their masses. We extend the binary demographics in M35 by many magnitudes faint-ward of previous (radial-velocity) studies of this cluster and further away from the cluster center ($1.78^{\circ}$). We find \binaries binaries down to Gaia $G = 20.3$ and determine a lower limit on the binary frequency of $\binfrac \pm \binfracsig $.  The binary fraction near the tidal radius is low and increases dramatically toward the cluster center (Figure~\ref{fig:fig2}).  

We find the binaries to be centrally concentrated as compared to the single stars, likely a result of both dynamical and primordial mass segregation (see Figures~\ref{fig:fig3} and \ref{fig:fig4}).  Investigating the M35 single and binary samples independently, we find strong evidence for mass segregation throughout the binary star population, while only the most massive singles appear to have begun moving towards a more centrally concentrated radial distribution. 

We estimate the half-mass relaxation time to be \trh $\pm$ $84$ Myr and use this to estimate mass segregation times for stars in the cluster based on their masses. Approximately $26\%$ of stellar systems in M35 have had time to become dynamically mass segregated. Thus, our results suggest that M35 is not yet fully relaxed. Interestingly, even for the stars that have mass segregation times greater than the cluster age, we still find the binaries to be significantly more centrally concentrated than the single stars.  As we do not expect these stars to have had time to become mass segregated dynamically given the current cluster structure, we conclude that the binaries became mass segregated during the cluster formation process, either by forming closer to the cluster center than the single stars and/or forming within smaller, more rapidly evolving, clumps that merged early on to form the cluster we observe today.  

This study is a precursor for future work investigating trends across many clusters, with specific emphasis on how binary characteristics evolve over time as a result of the local stellar environment.

\begin{acknowledgments}
This material is based upon work supported by the National Science Foundation (NSF) under grant No. AST-2149425, a Research Experiences for Undergraduates (REU) grant awarded to CIERA at Northwestern University, and under the NSF AAG Grant No. AST-2107738. Any opinions, findings, and conclusions or recommendations expressed in this material are those of the author(s) and do not necessarily reflect the views of the NSF. This research was supported in part through the computational resources and staff contributions provided for the Quest high performance computing facility at Northwestern University which is jointly supported by the Office of the Provost, the Office for Research, and Northwestern University Information Technology. A special thanks to Elizabeth Jefferey, Roger Cohen, Elliot Robinson, and the rest of the BASE-9 team.
\end{acknowledgments}




\bibliography{main}{}
\bibliographystyle{aasjournal}

\end{document}